\documentclass[pre,aps,twocolumn,showpacs,floatfix,10pt,reprint]{revtex4-1}
\usepackage[dvips]{graphicx}
\usepackage{amsmath}
\usepackage{amssymb}
\usepackage[nice]{nicefrac}
\usepackage{color}

\begin{document}

\title{Harmonic oscillator under L\'evy noise: Unexpected properties in the phase space}

\author{Igor M. Sokolov}
\email{igor.sokolov@physik.hu-berlin.de}
\affiliation{Institut f\"ur Physik, Humboldt-Universit\"at zu Berlin, Newtonstrasse 15, D--12489 Berlin, Germany}

\author{Bart{\l}omiej Dybiec}
\email{bartek@th.if.uj.edu.pl}
\affiliation{Marian Smoluchowski Institute of Physics, and Mark Kac Center for Complex Systems Research, Jagellonian University, ul. Reymonta 4, 30--059 Krak\'ow, Poland}

\author{Werner Ebelling}
\email{werner\_ebeling@web.de}
\affiliation{Institut f\"ur Physik, Humboldt-Universit\"at zu Berlin, Newtonstrasse 15, D--12489 Berlin, Germany}

\date{\today}
\begin{abstract}
A harmonic oscillator under influence of the noise is a basic model of
various physical phenomena.  Under Gaussian white noise the position
and velocity of the oscillator are independent random variables which
are distributed according to the bivariate Gaussian distribution with
elliptic level lines. The distribution of phase is homogeneous. None
of these properties hold in the general L\'evy case. Thus, the level
lines of the joint probability density are not elliptic. The
coordinate and the velocity of the oscillator are strongly dependent,
and this dependence is quantified by introducing the corresponding
parameter (``width deficit''). The distribution of the phase is
inhomogeneous and highly nontrivial.

\end{abstract}

\pacs{
 05.40.Fb, % Random walks and Levy flights
 05.10.Gg, % Stochastic analysis methods (Fokker--Planck, Langevin, etc.)
 02.50.-r, % Probability theory, stochastic processes, and statistics (see also section 05 Statistical physics, thermodynamics, and nonlinear dynamical systems)
 02.50.Ey, % Stochastic processes
 }
\maketitle

\section{Introduction}

Many situations in natural sciences can be successfully investigated
adopting the stochastic level of description, considering the system
of interest as a dynamical system responding to external perturbations
represented by a noise.  In the simplest situations this noise is
assumed to be white and Gaussian. The white type of the noise is the
consequence of the large number of independent interactions bounded in
time. Its Gaussian character arises due to the assumption that the
interactions are bounded in their strength.  In many cases  far from
equilibrium the second assumption fails; the noise still
can be considered as white but is now described by
heavy-tailed distributions, often of the $\alpha$-stable L\'evy
type \cite{janicki1994,*chechkin2006,*dubkov2008}.  Such heavy-tailed
fluctuations are abundant in turbulent fluid flows
\cite{shlesinger1993,*klafter1996,*solomon1993,
*delcastillonegrete1998}, magnetized plasmas \cite{chechkin2002b,
*delcastillonegrete2005}, optical lattices \cite{Katori1997},
heartbeat dynamics \cite{peng1993}, neural networks \cite{segev2002},
search on a folding polymers \cite{lomholt2005}, animal movement
\cite{Viswanathan1996}, climate dynamics \cite{ditlevsen1999b},
financial time series \cite{mantegna2000}, and even in spreading of
diseases and dispersal of banknotes \cite{brockmann2006}.  These
observations indicate the need for closer examination of properties of
stochastic systems under such heavy-tailed noises.

A damped harmonic oscillator under influence of external noise is a
basic model of non-equilibrium statistical physics. The evolution of
the state variable $x$ is described by the following equation of
motion
\begin{equation}
\ddot{x}(t)=-\gamma \dot{x}(t) - kx(t) + \xi(t)
\label{HaO}
\end{equation}
(the mass of the oscillating particle is taken to be unity).  Since
the response function of the harmonic oscillator is known, the
coordinate $x(t)$ and the velocity $v(t)= \dot{x}(t)$ are obtained by
the action of a known linear operators on the noise term $\xi(t)$, and
the characteristic function of the joint probability distribution of
$(x,v)$ can be obtained explicitly \cite{west1982} by generalizing the
results by Doob \cite{doob1942}.  However, although the formal
solution of the problem is known for almost 30 years, it doesn't seem
that the properties of such a basic system have attracted any
attention of physicists or mathematicians, although they are
astonishingly different from the ones under Gaussian noise. Here, we
focus on the properties of the corresponding stationary distribution,
which is achieved at times $t \gg \gamma/k$.

In the Gaussian case the distribution of
variables $(x,v)$ is a bivariate normal of the type
\begin{equation}
p(x,v)= \frac{1}{2\pi \sigma_x \sigma_v} \exp\left[- \frac{x^2}{2\sigma_x^2} - \frac{v^2}{2\sigma_v^2} \right]
\end{equation}
where $\sigma_x$ and $\sigma_v$ are the widths of the distributions of
the coordinate and velocity \cite{chanrasekhar1943}. Since this
distribution factorizes into a product of $x$- and $v$-distributions,
the phase variables are independent. The independence of $x$ and $v$
carries over to the equipartition theorem of equilibrium statistical
physics.  Since the argument of the exponential is a quadratic form of
the coordinate and velocity, the contours of equal probability density
are elliptic (the distribution is \textit{elliptically contoured}, or
simply \textit{elliptic}).  Introducing the rescaled variables $\tilde{x} =
x/\sigma_x$ and $\tilde{v} = v/\sigma_v$, one defines the phase angle
$\phi = \arctan(\tilde{v}/\tilde{x})$ which is uniformly distributed
over $[-\pi/2,\pi/2)$.

In the case of L\'evy noises other than the Gaussian one none of these
properties holds.  Thus, the variables $x$ and $v$ are dependent, and
the dependence is stronger in the overdamped case
($\sqrt{k}<\gamma/2$) than in the underdamped one
($\sqrt{k}>\gamma/2$). The distribution of $(x,v)$ is not elliptic
(see Fig.~\ref{fig:pdf2d}), and the phase angle shows an
inhomogeneous, highly nontrivial distribution (see
Fig.~\ref{fig:phase}). In what follows we present the results of
numerical simulations of the harmonic oscillator under L\'evy noise
together with analytical results corroborating these findings. We
moreover discuss in some detail the measures which can be used to
characterize dependence of the nonelliptic L\'evy variables and
quantify the dependence of $x$ and $v$-variables.

\section{General considerations}

Although the initial part of our calculations does not go beyond
\cite{west1982}, we reproduce part of the considerations in a physical
language for the sake of readability.  The formal solution of
Eq.~(\ref{HaO}) is
\begin{equation}
x(t) = F(t) + \int_{-\infty}^t G(t-t') \xi(t') dt',
\label{ForSol}
\end{equation}
where $G(t)$ is the Green's (response) function of the corresponding
process, and $F(t)$ is a decaying function (a solution of the
homogeneous equation under given initial conditions). The solution for
$v$ is given by
\begin{equation}
v(t) = F_v(t) + \int_{-\infty}^t G_v(t-t') \xi(t') dt',
\label{SolVel}
\end{equation}
where $G_v(t)$ is the Green's function of the velocity process
\begin{equation}
G_v(t)=\frac{d}{dt} G(t).
\end{equation}
In a stationary situation, $t \to \infty$, the $F$-functions in
Eqs.~(\ref{ForSol}) and (\ref{SolVel}) vanish. The Green's function of
Eq.~(\ref{HaO}) can be easily found e.g. via the Laplace
representation, and reads:
\begin{equation}
G(t)=\frac{\exp(-\gamma t/2)}{\sqrt{\omega^2-\gamma^2/4}} \sin\left[\sqrt{\omega^2-\gamma^2/4}\; t\right]
\end{equation}
for $\omega =\sqrt{k} > \gamma/2$ (underdamped case),
\begin{equation}
G(t)= t \exp(-\gamma t/2)
\end{equation}
for $\omega = \gamma/2$  (critical case)
and
\begin{equation}
G(t) = \frac{\exp(-\gamma t/2)}{\sqrt{\gamma^2/4-\omega^2}} \mathrm{sh} \left[\sqrt{\gamma^2/4-\omega^2}\;t\right]
\end{equation}
for $\omega < \gamma/2$ (overdamped case). Note that the functions $G(t)$ vanish both for $t=0$ and for
$t \to \infty$
so that
\begin{eqnarray}
\int_0^\infty G(t) \left[ \frac{d}{dt} G(t) \right] dt & =  & \frac{1}{2}\int_0^\infty \left[ \frac{d}{dt} G^2(t) \right] dt \\
& = & \frac{1}{2} \left. G^2(t) \right|_0^\infty =0, \nonumber
\end{eqnarray}
i.e. $G(t)$ and $G_v(t)$ are orthogonal on $[0,\infty)$.

\begin{figure}[!ht]
\begin{tabular}{p{4.0cm}p{4.0cm}}
\includegraphics[angle=0, width=4.0cm]{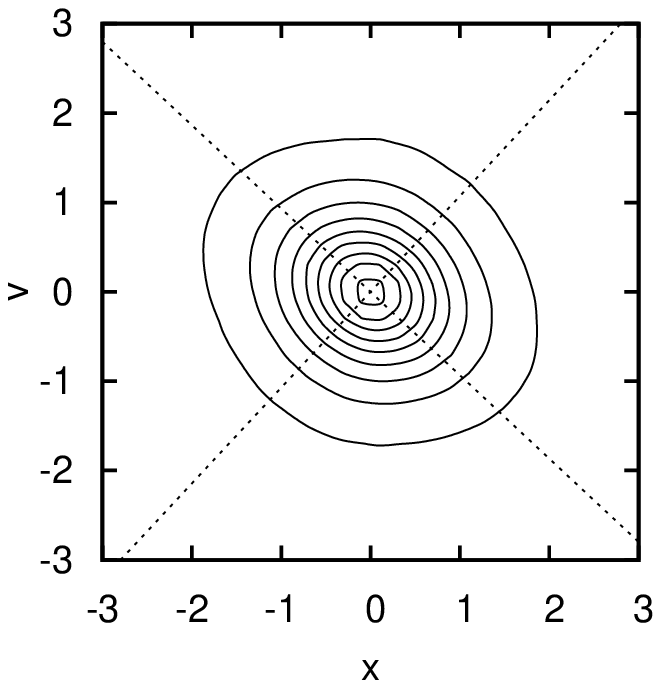} & \includegraphics[angle=0, width=4.0cm]{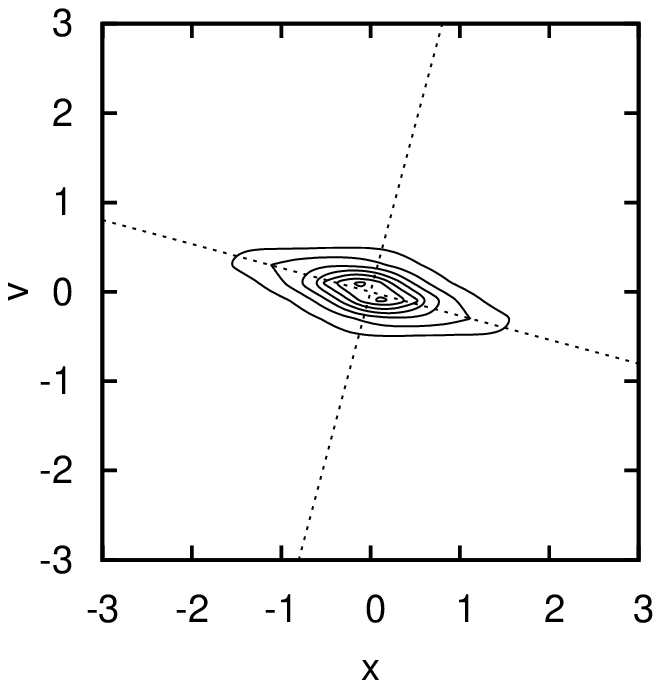}\\
\includegraphics[angle=0, width=4.0cm]{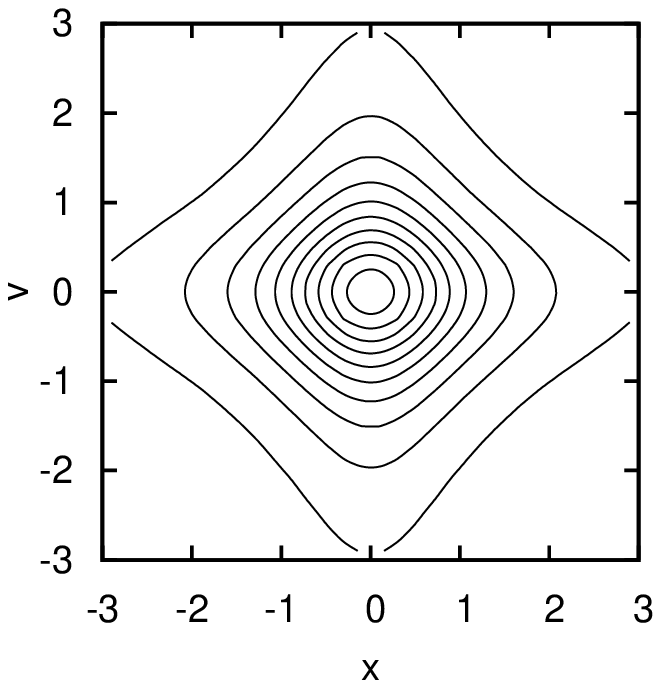} & \includegraphics[angle=0, width=4.0cm]{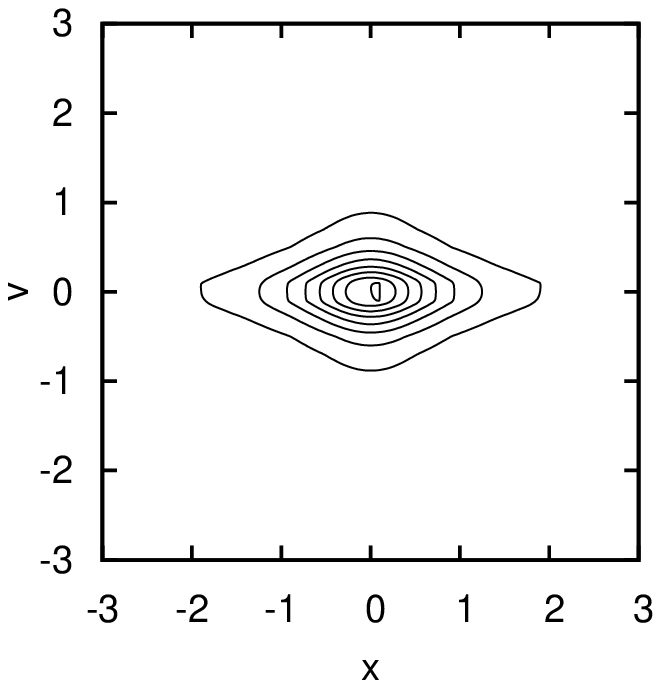}
\end{tabular}
\caption{Joint probability density $p(x,v)$ (top panel) and product of
two marginal densities $p_x(x) p_v(v)$ (bottom panel) for
$\alpha=1$. The left panel corresponds to the underdamped case
($k=1,\gamma=1$) while the right one to the overdamped case
($k=1,\gamma=4$). Dashed lines in the top panel represent major axes
described in the text. Simulation parameters: number of repetitions
$N=10^7$, time step of integration $\Delta t=10^{-2}$.}
\label{fig:pdf2d}
\end{figure}

Let us now confine ourselves to the finite (but long enough) interval
of integration and consider integral sums corresponding to
Eqs.~(\ref{ForSol}) and (\ref{SolVel}) for a given time step $\Delta
t$:
\begin{equation}
x = \sum_{i=1}^N a_i \xi_i
\end{equation}
and
\begin{equation}
v = \sum_{i=1}^N b_i \xi_i
\end{equation}
with $\xi_i$ being independent identically distributed random
variables corresponding to the integrals of the noise over independent
$\Delta t$-intervals, and $a_i$, $b_i$ being the values of the Green's
functions evaluated at corresponding points.

Let us evaluate the joint PDF of $x$ and $v$
\begin{eqnarray}
&& p(x,v)= \\
&& \!\!\!\!\! \int\dots\int \delta \left( x-\sum_{i=1}^N a_i \xi_i\right) \delta
\left( v-\sum_{i=1}^N b_i \xi_i\right)\prod_{i=1}^N p(\xi_i) d \xi_i. \nonumber
\end{eqnarray}
The characteristic function of this distribution is given by the Fourier transform
\begin{eqnarray}
f(k,q) & =  & \int\!\!\!\int e^{ikx+iqv} p(x,v) dx dv  \\
& = &  \prod_{i=1}^N \int \exp[(ika_i +iqb_i)\xi_i]p(\xi_i) d \xi_i \nonumber \\
& = & \prod_{i=1}^N f(ka_i + qb_i), \nonumber
\end{eqnarray}
where $f(k)= \int e^{ik\xi} p(\xi) d \xi$ is the characteristic
function of the distribution of $\xi$.

In the case considered here, the noise corresponds to a formal derivative of a
L\'evy random process $L(t)$. The integrals of the noise over finite
time intervals are independent L\'evy stable random variables.
The probability distributions corresponding to such random variables,
stable infinitely divisible laws, are limiting distributions in the space of probability
densities due to generalized central limit theorems
\cite{gnedenko1968,janicki1994}. The characteristic function of
$\xi(\Delta t)$ is thus given by
\begin{equation}
\langle e^{ik \xi(\Delta t)}\rangle = \exp\left[
-\Delta t\sigma^\alpha |k|^\alpha\left(1-i\beta\mathrm{sgn}
k\tan\frac{\pi\alpha}{2}\right) \right],
\label{eq:charact}
\end{equation}
where $\alpha\in(0,2]$ is the stability index, $\beta\in[-1,1]$ is the
asymmetry (skewness) parameter, while $\sigma$ ($\sigma>0$) is a scale
parameter.  Increments of the L\'evy motion are distributed according
to a $\alpha$-stable density, which for $\alpha<2$ has power law
asymptotics, i.e. $p(\xi) \propto 1/(\xi)^{\alpha+1}$. The
Gaussian distribution is a special case of $\alpha$ stable density
with $\alpha=2$, which is the only one $\alpha$-stable density
possessing finite moments \cite{janicki1994}. In our work here
we concentrate on the case of symmetric distributions, $\beta=0$.
The characteristic functions of symmetric L\'evy distributions have the form
\begin{equation}
f(k) = \exp(- \Delta t \sigma^\alpha |k|^\alpha).
\end{equation}
Therefore
\begin{equation}
\prod_{i=1}^N f(ka_i + qb_i) = \exp\left[- \Delta t \sigma^\alpha \sum_{i=1}^N |ka_i + qb_i|^\alpha \right],
\end{equation}
or returning to the integral notation,
\begin{equation}
f(k,q) = \exp\left[- \sigma^\alpha \int_0^{\infty} |k G(t) + q G_v(t)|^\alpha dt \right].
\label{ChaF}
\end{equation}
The characteristic functions of the marginal distributions are
obtained by putting $q=0$ (for $x$-distribution) or $k=0$ (for
$v$-distribution). Note moreover, that the characteristic function of
the distribution of any linear combination of $x$ and $v$
\begin{equation}
z=ax+bv
\end{equation}
will read
\begin{equation}
f_z(k) = f(a k, b k)
\label{eq:generalcf}
\end{equation}
and is the one of the symmetric L\'evy distribution,
so that the distribution of $x$ and $v$ is a \textit{bona fide} bivariate stable one.
It is however not of elliptic type, for which case the characteristic function would be
\begin{equation}
f(k,q) =  \exp\left[- \sigma^\alpha \int_0^{\infty} |w_x^2 k^2 + C_{xv} kq +w_v^2 q^2|^{\alpha/2} dt \right]
\end{equation}
i.e. a function of a quadratic form in $k$ and $q$, see \cite{press1972}.

The characteristic function, Eq.~(\ref{ChaF}) does not decouple into the product of the
functions of $k$ and $q$ only,
i.e. the distributions of $x$ and $v$ are \textit{not independent} (we shall say, they are \textit{associated}),
except for the Gaussian ($\alpha = 2$) case. In this case
\begin{eqnarray}
\int_0^{\infty} |k G(t) + q G_v(t)|^2 dt & = & k^2 \int_0^\infty G^2(t) dt \\
& + &  q^2 \int_0^\infty G_v^2(t) dt \nonumber \\
& + & 2 kq \int_0^\infty G(t)G_v(t) dt, \nonumber
\end{eqnarray}
and the last integral vanishes due to orthogonality of $G(t)$ and $G_v(t)$. In all other cases such decoupling
does not take place. The Gaussian case is also the only case when the distribution is elliptic.

\begin{figure}[!ht]
\includegraphics[angle=0, width=8.0cm]{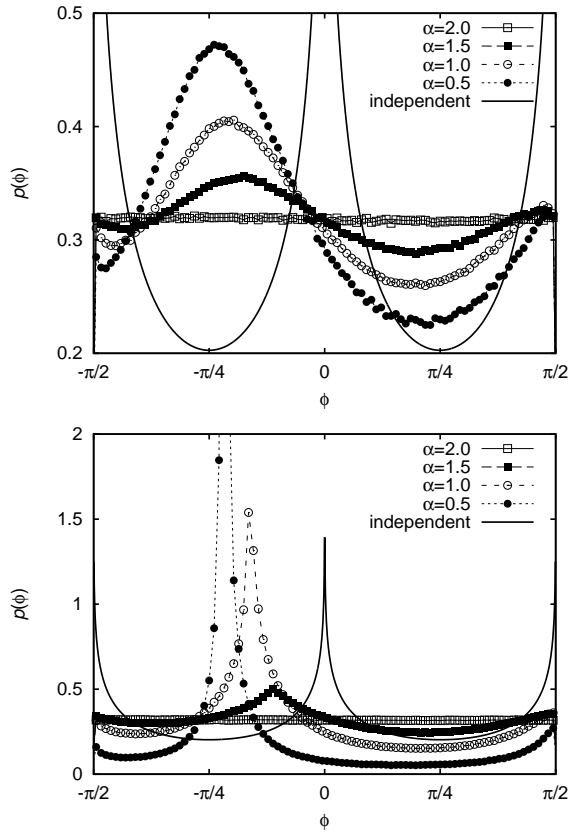}
\caption{Phase distribution $p(\phi)$ in the underdamped
($k=1,\gamma=1$ -- top panel) and overdamped ($k=1,\gamma=4$ -- bottom
panel) cases for different values of $\alpha$. The full line shows the
corresponding distribution for independent Cauchy variables, see Eq.~(\ref{eq:independentcauchy}).
Simulation parameters as in Fig.~\ref{fig:pdf2d}}
\label{fig:phase}
\end{figure}

In order to demonstrate lack of independence of $x$ and $v$ for the harmonic oscillator under L\'evy noise
we have estimated from numerical simulations of the process the joint distribution $p(x,v)$ and compared it with product
of two marginal densities $p_x(x) p_v(v)$, see Fig.~\ref{fig:pdf2d}, presented
both for the underdamped (left panel) and for the
overdamped (right panel) case under Cauchy ($\alpha=1$) noise \cite{garbaczewski2000}.
The contours of the corresponding joint probability densities are
central symmetric, and the positions of their main axes, along which the width is the largest or
the smallest, can be obtained numerically by finding the extrema of
\begin{equation}
w(\theta)= \int_0^\infty  |G(t) \cos \theta + G_v(t)\sin \theta|^\alpha dt.
\end{equation}
The corresponding axes are at the angles of approximately $-43^\circ$
(longer axis) and $47^\circ$ (shorter one) for the underdamped case,
and of approximately $-15^\circ$ (longer axis) and $75^\circ$ (shorter
one) for the overdamped case.  The corresponding angles are close (but
not equal to) the positions of extrema in the phase plot, due to
different weighting of the variables, compare Figs.~\ref{fig:pdf2d}
and \ref{fig:phase}.

\section{Measures of dependence}

The case of dependent random variables lacking moments poses a problem of quantifying
the strength of their dependence. In the ``normal'' Gaussian case, and in every case when
the second moments of the variables are finite, the coefficient of covariance gives us a
standard measure of such dependence. For variables with zero mean
\begin{equation}
\mathrm{cov}(x,y) = \langle x  y \rangle,
\end{equation}
and for variables whose mean values are nonzero the same is defined for the centered variables
$x_c=x-\bar{x}$ and $y_c= y-\bar{y}$. In what follows we consider only variables centered at zero, and omit the subscript $c$.
To characterize the relative strength of the dependence
one can introduce the correlation coefficient $\mathrm{corr}(x,y)$ by normalizing the covariance
over the dispersions of the corresponding centered variables. Alternatively, one can consider the
covariance of variables $x$ and $y$ normalized with respect to their dispersions
$
\sigma_x = \langle \tilde{x}^2 \rangle^{1/2}
$
and
$
\sigma_y = \langle \tilde{y}^2 \rangle^{1/2},
$
i.e.
$\tilde{x} =x/\sigma_x$, $\tilde{y} =y/\sigma_y$:
\begin{equation}
\mathrm{corr}(x,y) = \frac{\mathrm{cov}(x,y)}{\sigma_x \sigma_y} = \mathrm{cov}(\tilde{x},\tilde{y}).
\end{equation}
The correlation coefficient of two random variables changes in the interval $[-1,1]$ and it is unity if the
variable $y$ is a copy of $x$ up to arbitrary rescaling, and to minus unity if $y$ is a copy of $x$ taken with the opposite sign,
again up to the arbitrary change of scale.
The correlation coefficient of independent variables vanishes. Since the definition of the
correlation coefficient involves the second moment of the corresponding random variables, it cannot be
used immediately for the ones lacking such a moment. Here several generalizations have been proposed,
all seeming to be deficient for our purpose, consequently we have to extend their definitions.

A relatively common measure of dependence of L\'evy variables is the \textit{codifference} \cite{samorodnitsky1994}.
To understand its nature it is enough first to consider the Gaussian case and to turn to the
characteristic function of the corresponding bivariate distribution of $x$ and $y$
\begin{eqnarray}
f(k_1, k_2) &=&\int\int dx dy e^{ik_1 x + ik_2 y} p(x,y) \\
&=& \exp \left[-\sigma_x^2 k_1^2 -\sigma_y^2 k_2^2 - 2\mathrm{cov}(x,y)k_1k_2\right]. \nonumber
\end{eqnarray}
To assess the covariance one can then consider the difference variable $d=x-y$ whose
characteristic function is
\begin{eqnarray}
 f_d(k) & = & f(k,-k) \\ \nonumber
& =  & \exp \left[-\left(\sigma_x^2 +\sigma_y^2 -2\mathrm{cov}(x,y)\right)k^2\right],
\end{eqnarray}
from which
\begin{equation}
\mathrm{cov}(x,y) =\frac{1}{2} \left[ \ln \left. f_d(k) \right|_{k=1} + \sigma_x^2 +\sigma_x^2 \right],
\end{equation}
or in slightly different notation
\begin{equation}
\mathrm{cov}(x,y) = \frac{1}{2} \mathrm{codiff}(x,y)
\end{equation}
with $\mathrm{codiff}(x,y)$
being the codifference of the variables $x$ and $y$ defined through
\begin{eqnarray}
\mathrm{codiff}(x,y) & = & \ln \left. f(k,-k) \right|_{k=1} \\ \nonumber && + \ln \left. f(k,0) \right|_{k=1} + \ln \left. f(0,k) \right|_{k=1}.
\end{eqnarray}
The definition of the codifference as a measure of dependence via characteristic functions does not rely on the
existence of any moments and is universal, i.e. it can be used in the L\'evy case as well. We note that equivalently
one can use the sum of the two variables $x$ and $y$, and to define the ``cosum'' as
\begin{eqnarray}
\mathrm{cosum}(x,y) & =  & -\ln \left. f(k,k) \right|_{k=1} \\ \nonumber & & + \ln \left. f(k,0) \right|_{k=1} + \ln \left. f(0,k) \right|_{k=1}
\end{eqnarray}
The codifference (or cosum) as a measure of dependence has two important drawbacks, the first one quite specific to their definitions,
and the second one inherited from the the definition of covariance itself.

First, defining the corresponding parameters via characteristic functions, being the
means of strongly oscillating exponentials implies numerically generating very large data samples.
Therefore, for the sake of practicability, the measure of dependence has to rely on some robust
and easily accessible statistics.

Second, using the codifference
as a measure of dependence of the observables of different dimension (coordinate and velocity) makes no sense
physically, since the change in units of measurement changes the measure of dependence.
The codifference measure is often used in the analysis
of time series, since the dimensions of all terms are the same, the problem
of units therefore do not stand. The same is true also for the normalized codifference \cite{rosadi2007}, which is to
no extent a simple analogue of the correlation coefficient.

Let us first address the problem of statistics. For a multivariate L\'evy distributions, as the one appearing in the previous Section, both the distributions
of the the difference
$
d=x-y
$, $p_d(d)$,
and of the sum
$
 s=x+y
$, $p_s(s)$,
of $x$ and $y$ (in our initial problem
the variable $y$ corresponds to the velocity $v$ of the particle) are univariate L\'evy distributions.
The widths of these distributions are exactly the prefactors of $|k|^\alpha$ in the corresponding characteristic functions
\begin{equation}
w_d^\alpha =  |\ln \left. f(k,-k)\right|_{k=1} |
\end{equation}
and
\begin{equation}
w_s^\alpha =  |\ln \left. f(k,k)\right|_{k=1}|.
\end{equation}
Analogously $w_x$ and $w_y$ are  equal to
\begin{equation}
w_x^\alpha=|\ln \left. f(k,0) \right|_{k=1}|
\end{equation}
 and
\begin{equation}
w_y^\alpha=|\ln \left. f(0,k) \right|_{k=1}|.
\end{equation}
Therefore
\begin{equation}
\mathrm{codiff}(x,y) = w_d^\alpha - w_x^\alpha -w_y^\alpha,
\end{equation}
and
\begin{equation}
\mathrm{cosum}(x,y) = w_s^\alpha - w_x^\alpha -w_y^\alpha.
\end{equation}
% where
% \begin{equation}
% w_x= \sigma \left[\int_0^{\infty} |G(t)|^\alpha dt \right]^{1/\alpha}
% \end{equation}
%  and
% \begin{equation}
% w_v= \sigma \left[\int_0^{\infty} |G_v(t)|^\alpha dt \right]^{1/\alpha}.
% \end{equation}

Contrary to the initial definition via characteristic functions, the ones over the widths (scaling
parameters) of the distributions can be easily assessed numerically since they are connected with the
robust characteristics like the interquartile distances of the corresponding distributions or with the
heights of their probability densities at zero
\begin{equation}
p(0) = \frac{1}{2\pi} \int_{-\infty}^{\infty} e^{-w^\alpha |k|^\alpha} dk = \frac{\Gamma(1/\alpha)}{\pi w},
\end{equation}
which are easier to read out from the numerically obtained histograms. This ``height measure''
is repeatedly used in our work.

To circumvent the second problem, we propose the following measure of dependence $D$
corresponding to the cosum of the normalized variables.
Both distributions, the one of $x$ and the one of $v$
are first normalized by introducing $\tilde{x} = x/w_x$ and $\tilde{v} = v/w_y$ (both possessing
L\'evy distributions of unit width). Then we compare the width of the distribution of $\zeta = \tilde{x} + \tilde{v}$
(raised to the power of $\alpha$) with 2, the sum of width of the distributions of $\tilde{x}$ and $\tilde{v}$:
\begin{equation}
D = \frac{w_\zeta^\alpha - 2}{2},
\label{eq:widthexcessdefinition}
\end{equation}
and call this measure of dependence the width excess (the letter $D$ stands here for ``Dependence'').
The measure $D$, the cosum of rescaled variables,
changes in the interval $[-1,1]$. It is zero for independent variables,
equal to 1 if the variable $v$ is a copy of $x$ up to arbitrary rescaling (i.e. up to the change
of units) and is equal to $-1$ if the variable $v$ is a copy of $-x$, up to rescaling.
It is clear that $D$ is insensitive to the units in which $x$ and $v$ are measured.

For a bivariate Gaussian density $D$ is exactly the correlation
coefficient
\begin{equation}
 D = \frac{\left\langle xv \right\rangle}{ \sigma_x \sigma_v}=\mathrm{corr}(\tilde{x},\tilde{y}).
\end{equation}
For the elliptic L\'evy process it is connected with the Press'
association parameter \cite{press1972}, which however can hardly be
generalized to nonelliptic cases and vanishes for isotropic
distributions, the ones with characteristic function
\cite{samorodnitsky1994}
\begin{equation}
f(k,q)=\exp\left[-\sigma^\alpha (k^2+q^2)^{\alpha/2}\right],
\label{eq:isotropic}
\end{equation}
although the corresponding $x$ and $v$ values are not independent, see
Eqs.~(\ref{eq:widthexcessdefinition}) and ~(\ref{eq:isotropicvalues}).

The characteristic function of the distribution of $\zeta$
\[
\zeta = \tilde{x} + \tilde{v}=x/w_x+v/w_v
\]
is
\[
f_\zeta(k) = f(k/w_x, k/w_v).
\]
Consequently the width excess is
\begin{equation}
D = \frac{1}{2}\int_0^{\infty} \left| \frac{G(t)}{I_1} + \frac{G_v(t)}{I_2}\right|^\alpha dt  -1
\label{eq:widthexcess}
\end{equation}
with
\begin{equation}
I_1 = \left[\int_0^{\infty} |G(t)|^\alpha dt \right]^{1/\alpha}
\end{equation}
and
\begin{equation}
I_2 = \left[\int_0^{\infty} |G_v(t)|^\alpha dt \right]^{1/\alpha}.
\end{equation}

\begin{figure}[!ht]
\includegraphics[angle=0, width=8.0cm]{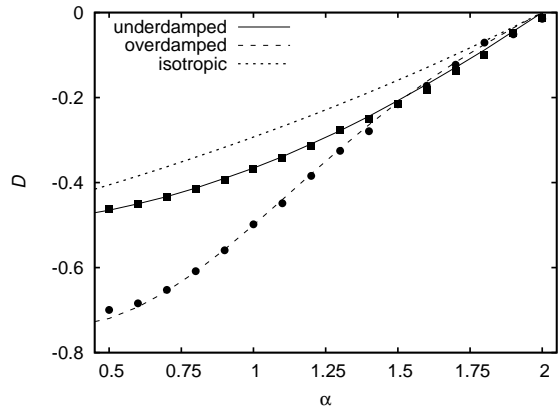}
\caption{Width excess $D$, see Eq.~(\ref{eq:widthexcess}), for various
$\alpha$. Lines represent theoretical formulas for underdamped
($k=1,\gamma=1$) and overdamped ($k=1,\gamma=4$) cases, points
correspond to simulation results. The dashed line presents values of the width excess $D$ for isotropic case, see Eqs.~(\ref{eq:isotropic}) and~(\ref{eq:isotropicvalues}).
Other simulation parameters as in
Fig.~\ref{fig:pdf2d}.
}
\label{fig:d}
\end{figure}

The width excess (\ref{eq:widthexcess}) vanishes for $\alpha=2$ only, indicating the
independence of the position and velocity variables. Contrary to the $\alpha=2$ case, for
$\alpha<2$ it takes negative values, indicating anti-association, i.e. dependence between the position and the velocity.
The results of simulations and calculations of the width deficit for
the underdamped and overdamped cases are given in Fig. \ref{fig:d},
together with the result for an isotropic case, for which
\begin{equation}
D=2^{\alpha/2-1}-1
\label{eq:isotropicvalues}
\end{equation}
corresponding to all elliptic distributions, see
Eqs.~(\ref{eq:widthexcessdefinition}) and (\ref{eq:isotropic}).
Figure~\ref{fig:d} demonstrates perfect agreement between theoretical
results given by formula~(\ref{eq:widthexcess}) (lines) and numerical data (points).  Negative values of the
width excess $D$ indicate strong negative association, which is
stronger for the overdamped case ($k=1,\gamma=4$) than for the
underdamped case ($k=1,\gamma=1$).

\section{Phase and amplitude distributions}

The strong anti-association of velocity and coordinate is reflected in
the distribution of the phase angle
\begin{equation}
\phi=\arctan\left[ \frac{\tilde{v}}{\tilde{x}} \right]=\arctan\left[ \frac{v/w_v}{x/w_x} \right].
\end{equation}
This distribution is homogeneous over $[-\pi/2,\pi/2)$ for all
elliptic bivariate L\'evy laws (e.g. in the Gaussian case) and has a
marked form with peaks (divergences) at $\phi = 0, \pm \pi/2$ for
independent random L\'evy variables (corresponding to the star-like
form of the product of two L\'evy distributions, see
Fig. \ref{fig:pdf2d}).  The actual forms of the phase PDFs are shown
in Fig.~\ref{fig:phase} for the underdamped (top panel) and overdamped
(bottom panel) cases.
It is interesting that the peaks are present at ``nontrivial'' angles,
the ones different from $0$ and $\pm \pi/2$.

\begin{figure}[!ht]
\includegraphics[angle=0, width=8.0cm]{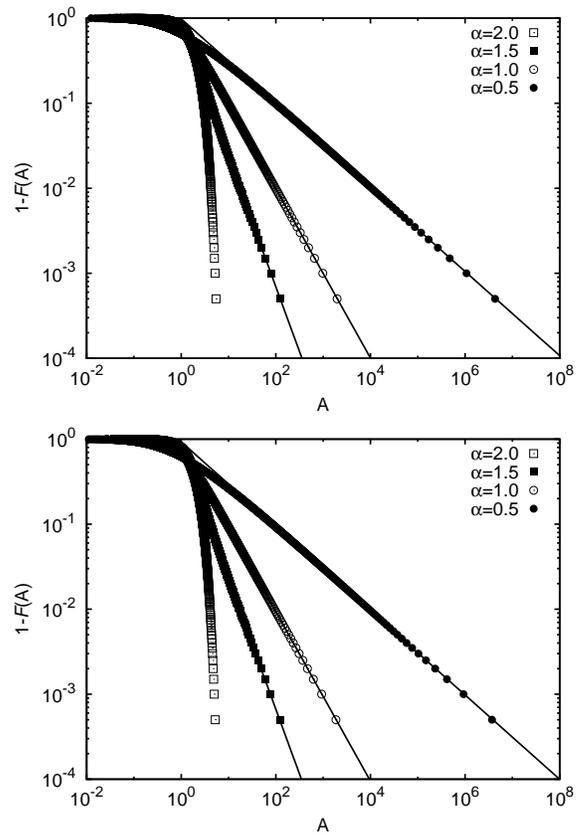}
\caption{Complementary cumulative distribution ($1-F(A)$) of the
amplitude $A=\sqrt{\tilde{x}^2+\tilde{v}^2}=\sqrt{(x/w_x)^2+(v/w_v)^2}$ in the underdamped ($k=1,\gamma=1$ --
top panel) and overdamped ($k=1,\gamma=4$ -- bottom panel) cases for
different values of $\alpha$. Full lines show the corresponding
asymptotic power-law decays of complementary cumulative distributions,
i.e. $A^{-\alpha}$.  Simulation parameters as in Fig.~\ref{fig:pdf2d}}
\label{fig:amplitude}
\end{figure}

For $\alpha<2$, strong deviations both from the uniform distribution and from the
product distribution are well visible. For $\alpha=1$ the phase distribution for independent $x$- and $v$-variables is given by
\begin{equation}
p(\phi)=\frac{1}{\pi^2 \cos 2 \phi} \ln\left[ \frac{1+\cos 2 \phi}{1-\cos 2 \phi}\right],
\label{eq:independentcauchy}
\end{equation}
see Fig.~\ref{fig:phase}.

The amplitude
\begin{equation}
A=\sqrt{\tilde{x}^2+\tilde{v}^2}=\sqrt{(x/w_x)^2+(v/w_v)^2}
\end{equation}
 is asymptotically distributed
according to a power-law distribution
\begin{equation}
p(A) \propto \frac{1}{A^{\alpha+1}},
\end{equation}
consequently the complementary cumulative distribution has also
power-law tails which are characterized by the exponent $\alpha$, see
Fig.~\ref{fig:amplitude}, as it follows from the results of Ref.~\cite{west1982}.

\section{Summary}

In situations far from equilibrium interactions of the harmonic
oscillator with the environment can often be described by the white
L\'evy noise.  Presence of L\'evy noises introduces dependence
(association) between velocity and position, which vanishes only in
the Gaussian limit of $\alpha=2$. The presence of association between
position and velocity is manifested by the nontrivial joint
distribution of $p(x,v)$ and nontrivial phase distribution $p(\phi)$.
Our main finding corresponds to strong anti-association between the position
and the velocity of a L\'evy-driven oscillator, a property which might
be of high importance for first passage properties of such a process.
For example, in the Kramers problem the dependence of position and velocity
might be responsible for the breakdown of the transition-state description.
A direct consequence of (anti-)association between velocity and position
is the breakdown of equipartition theorem of equilibrium statistical
physics for the harmonic L\'evy oscillator. The recorded findings,
together with results of earlier investigations
\cite{chechkin2003,chechkin2004,chechkin2006} clearly show that
properties of out-of-equilibrium (L\'evy noise driven) systems are
very different from their equilibrium (Gaussian noise driven)
counterparts.

%%%%%%%%%%%%%%%%%%%%%%%%%%%%%%%%%%%%%%%%%%%%%%%%%%%%%%%%%%%
%%%%%%%%%%%%%%%%%%%%%%%%%%%%%%%%%%%%%%%%%%%%%%%%%%%%%%%%%%%
%
% BIBLIOGRAPHY
%
%%%%%%%%%%%%%%%%%%%%%%%%%%%%%%%%%%%%%%%%%%%%%%%%%%%%%%%%%%%
%%%%%%%%%%%%%%%%%%%%%%%%%%%%%%%%%%%%%%%%%%%%%%%%%%%%%%%%%%%
% \bibliography{bibliography}

%Merlin.mbs v4.21 2009-07-09.
%

\end{document}